\documentstyle[prb,aps,eqsecnum,twocolumn]{revtex}
\begin{document}
\draft
\twocolumn[\hsize\textwidth\columnwidth\hsize\csname@twocolumnfalse\endcsname
\title{
Magnetic and charge structures in itinerant-electron magnets:\\
Coexistence of multiple SDW and CDW}
\author{Fusayoshi J. Ohkawa}
\address{Division of Physics, Graduate School of Science, 
Hokkaido University, Sapporo 060-0810, Japan}
\date{Received: \hspace{4cm} }
\maketitle
\begin{abstract}
A theory of Kondo lattices is applied to studying possible
magnetic and charge structures of itinerant-electron
antiferromagnets. Even helical spin structures can be stabilized
when the nesting of the Fermi surface is not sharp and the
superexchange interaction, which arises from the virtual exchange
of pair excitations across the Mott-Hubbard gap, is mainly
responsible for magnetic instability. Sinusoidal spin structures
or spin density waves (SDW) are only stabilized when the nesting 
of the Fermi surface is sharp enough and  a novel exchange
interaction arising from that of pair excitations of
quasi-particles is mainly responsible for magnetic instability. 
In particular, multiple SDW are stabilized when their
incommensurate ordering wave-numbers $\pm{\bf Q}$ are multiple; 
magnetizations of different $\pm{\bf Q}$ components are
orthogonal to each other in double and triple SDW when magnetic
anisotropy is weak enough. Unless $\pm 2{\bf Q}$ are
commensurate, charge density waves (CDW) with $\pm 2{\bf Q}$
coexist with SDW with $\pm{\bf Q}$. Because the quenching of
magnetic moments by the Kondo effect depends on local numbers of
electrons, the phase of CDW or electron densities is such that 
magnetic moments are large where the quenching is weak. It is
proposed that the so called stipe order in cuprate-oxide
high-temperature superconductors must be the coexisting state of
double incommensurate SDW and CDW. 
\end{abstract}
%
~\\]
%

\section{Introduction}\label{SecIntroduction}

It is a long standing and important issue to explain two types
of magnetism, local-moment magnetism and itinerant-electron
magnetism, in a unified theoretical framework.   
Some of their physical properties are similar to each other and
others are different from each other.
Even if they are phenomenologically similar to each other, their
microscopic explanations can be totally different from each
other. For example, the spin susceptibility obeys the
Curie-Weiss law in both types of magnets. The Curie-Weiss law in
insulators is due to  the temperature dependence  of local
thermal spin fluctuations. This mechanism is of leading order in
$1/d$, with $d$  spatial dimensionality. On the other hand, the
Curie-Weiss law in metals is a controversial issue.  Two
mechanisms were proposed: the temperature dependence of the 
mode-mode coupling between intersite spin
fluctuations\cite{Murata,MK,Moriya} and that of Weiss' magnetic
mean fields.\cite{OhSDW-CW,Miyai} The former is of higher order
in $1/d$, while the latter is of leading order in $1/d$. It is
interesting which is responsible for the Curie-Weiss law in
actual $d=3$ dimensions. There is a sharp peak at the chemical
potential in the density of states in typical ferromagnetic
metals, and there is sharp nesting of the Fermi surface in
typical antiferromagnetic metals. Recent
theories\cite{Miyai,Miyake} showed that in such 
itinerant-electron magnets the mode-mode coupling plays a
totally negative role in the Curie-Weiss law. On the other hand,
a novel exchange interaction arising from the virtual exchange
of pair excitations of quasi-particles can have a large
temperature dependence consistent with the Curie-Weiss
law;\cite{OhSDW-CW,Miyai}  Weiss' mean fields are given by
magnetic polarizations multiplied by the exchange interaction.

Magnetic structures are different between the two types of
antiferromagnets. Helical structures are stabilized in typical
local-moment magnets,\cite{Yoshimori} and sinusoidal structures
or spin density waves (SDW)  in typical itinerant-electron
magnets. When ordering wave-numbers are incommensurate, there
are two or more than two equivalent wave-numbers depending on
lattice symmetry. An interesting issue is which are stabilized,
single or multiple SDW, in what conditions.

Itinerant-electron magnets lie in the vicinity of the
Mott-Hubbard metal-insulator transition.  A combined theory of
Hubbard's\cite{Hubbard} and Gutzwiller's\cite{Gutzwiller}
implies that in metallic phases the density of states is of a
three-peak structure, Gutzwiller's quasi-particle band at the
chemical potential between the lower and upper Hubbard
bands lying far from the chemical potential.\cite{ComThreePeak}
It is unquestionable that the Mott-Hubbard splitting exists even
in metallic phases in the vicinity of the transition. Such metals
must show a crossover between the two types of magnetism as a
function of temperatures. Denote the energy scale of local
quantum spin fluctuations by $k_{B}T_{K}$. They must behave as
local-moment magnets at $T\gg T_{K}$ because local thermal spin
fluctuations are dominant, and they behave as
itinerant-electron magnets at $T\ll T_{K}$ because magnetic
moments are quenched by local quantum spin fluctuations.  When
physical phenomana relevant to the Mott-Hubbard transition
are examined,  first of all, local spin fluctuations should be
accurately taken into account.

Local spin fluctuations are rigorously considered  in any
single-site approximation (SSA) that includes all the
single-site terms. Such an SSA is reduced to determining and
solving selfconsistently the Anderson model,\cite{Mapping}
which is one of the effective Hamiltonians for the Kondo
problem.  Because the Kondo problem was
solved,\cite{Wilson,Nozieres,Yamada,Yosida,Exact} many useful
results are available in clarifying single-site properties in
lattice systems. The three-peak structure discussed above can
be easily understood by the SSA theory or the mapping to the
Anderson model. The quenching of magnetic moments by local
quantum spin fluctuations in lattice systems is mapped to that
in the Kondo problem, so that $T_{K}$ discussed above is
nothing but the Kondo temperature.  A perturbative treatment of
intersite terms starting from an {\it unperturbed} state
constructed in the SSA is nothing but a theory of Kondo
lattices. Because the SSA is rigorous for Landau's normal
Fermi-liquid states in infinite dimensions
$(d\rightarrow+\infty)$,\cite{Metzner} it can also be formulated
as a $1/d$ expansion theory.

One of the purposes of this paper is to study which are
stabilized in itinerant-electron magnets, single or multiple
${\bf Q}$ structures, in what conditions. This paper is organized
as follows:  In Sec.~\ref{SecFormulation}, the theory of Kondo
lattices is applied to deriving Landau's free energy. What are
studied in Sec.~\ref{MagStr} are magnetic and charge structures
of ordered states below the N\'{e}el temperature. Discussion is
given in Sec.~\ref{Discussion}, and conclusion is given in
Sec.~\ref{Conclusion}. Landau's free energy for the Anderson
model is studied in Appendix~\ref{Anderson}.

\section{Formulation}\label{SecFormulation}
\subsection{Theory of Kondo lattices}\label{SecKondoTh}

Consider the following single-band model:
\begin{eqnarray}\label{EqHam}
{\cal H} &=& -  \! \sum_{ij \sigma} t_{ij}
d_{i \sigma}^{\dagger} d_{j \sigma} +
\! \sum_{i} U \hat{n}_{i \uparrow} \hat{n}_{i \downarrow} 
\nonumber \\ &&
- \frac{1}{2} \sum_{i \ne j} J_{ij}^{(p)}
\left( \hat{\bf s}_{i} \cdot \hat{\bf s}_{j} \right)
+ \frac{1}{2} \sum_{i \ne j} V_{ij}^{(c)}
\hat{n}_{i} \hat{n}_{i}  ,
\end{eqnarray}
with $\hat{n}_{i\sigma}= d_{i\sigma}^{\dagger}d_{i\sigma}$ and
$\hat{n}_{i}= \hat{n}_{i\uparrow} + \hat{n}_{i\downarrow}$;
$\hat{s}_{i\nu} = \frac{1}{2} \sum_{\alpha\beta}  
d_{i\alpha}^\dagger \sigma_\nu^{\alpha\beta} d_{i\beta}$ 
($\nu=x$, $y$ and $z$), with $\sigma_{\nu}^{\alpha\beta}$ the
Pauli matrix.  When only the first and second terms are
considered, the model (\ref{EqHam}) is reduced to the Hubbard
model, in which the wellknown superexchange interaction is
involved. Actual systems must be described by a multi-band
model. A superexchange interaction arising from multi-band
effects\cite{Roth,Multi-band} is phenomenologically included by
the third term. The fourth term is  the long range  Coulomb
interaction:
$V_{ij}^{(c)} = e^2/|{\bf R}_{i} - {\bf R}_{j}|$, with 
${\bf R}_{i}$ lattice sites.  No spin-orbit interaction or no
magnetic anisotropy is included in this paper.

Assume  Landau's normal Fermi-liquid state. The single-particle
selfenergy is divided into single-site and multi-site terms: 
$\Sigma_\sigma(i\varepsilon_n,{\bf k})=\tilde{\Sigma}_\sigma
(i\varepsilon_n)+\Delta\Sigma_\sigma(i\varepsilon_n,{\bf k})$. 
The multi-site term,
$\Delta\Sigma_\sigma(i\varepsilon_n,{\bf k})$,   is of higher
order in $1/d$, and is ignored here. The single-site term, 
$\tilde{\Sigma}_\sigma (i\varepsilon_n)$, is of leading order in
$1/d$. In the presence of infinitesimally small
external fields,
${\cal H}^\prime_{ext} = -\sum_{i\sigma} \bigl[\Delta\mu^\prime  +
\frac{1}{2}\sigma g\mu_B H^\prime \bigr]\hat{n}_{i\sigma}$, 
it is expanded in such a way that
$\tilde{\Sigma}_\sigma (i \varepsilon_n) =
\tilde{\Sigma}_0 + (1- \tilde{\phi}_\gamma) i \varepsilon_n +
(1-\tilde{\phi}_s) \frac{1}{2} \sigma g\mu_B  H^\prime   +
(1-\tilde{\phi}_c) \Delta\mu^\prime + \cdots$ for
$|\varepsilon_n| \ll k_BT_K$ and $T \ll T_K$, with $T_K$ 
the Kondo temperature discussed in Introduction. Because 
$\tilde{\Sigma}_\sigma(i\varepsilon_n)$ is equal to that of
the mapped Anderson model, it follows according to Yamada and
Yosida's\cite{Yamada,Yosida} that
$\tilde{\phi}_\gamma = (\tilde{\phi}_s + \tilde{\phi}_c)/2$;
$\tilde{\phi}_s$ and $\tilde{\phi}_c$  are approximately given by
Eqs.~(\ref{EqTilPhiS}) and (\ref{EqTilPhiC}) in Appendix. 
The electron density is given by\cite{ComNeutraility}
\begin{equation}\label{EqND}
n_{d} = \frac{1}{N} \sum_{i} \langle \hat{n}_i \rangle.
\end{equation}
with $N$ the total number of unit cells. When the system is in
the vicinity of the Mott-Hubbard transition, $n_{d}$ is close
to unity so that 
$\tilde{\phi}_s \gg 1$ and $\tilde{\phi}_c \ll 1$. 
The single-particle Green function is given by
\begin{equation}\label{EqGreen}
G_\sigma (i\varepsilon_n,{\bf k}) =
\frac{1}{\tilde{\phi}_\gamma}
\frac{1}{i\varepsilon_n - \xi({\bf k})},
\end{equation}
for $|\varepsilon_n| \ll k_B T_K$ and $T\ll T_K$, with
\begin{equation}
\xi({\bf k}) = \frac{1}{\tilde{\phi}_\gamma}
\biggl[\tilde{\Sigma}_0  - \sum_{j} t_{ij}e^{i {\bf k} \cdot 
({\bf R}_i-{\bf R}_j)} - \mu \biggr] .
\end{equation}
The quasi-particle density of states is given by
$\rho^*(\epsilon)= (1/N)\sum_{\bf k}\delta(\epsilon-\xi({\bf
k}))$. When  contributions from local spin fluctuations are only
considered, the specific heat at $T \ll T_K$ is given by
$\tilde{C}=\tilde{\gamma} T + \cdots$, with
$\gamma = (2/3)\pi^2 k_B^2  \rho^*(0)$.

When irreducible two-point polarization functions in the
wave-number representation are denoted by
$\pi_{\sigma\tau}(i\omega_{l},{\bf q})$, those in spin and charge
channels are given by
%
$\pi_{s}(i \omega_{l}, {\bf q})=
\pi_{\sigma\sigma}(i \omega_{l}, {\bf q})-
\pi_{\sigma-\sigma}(i \omega_{l}, {\bf q}) $
%
and
%
$\pi_{c}(i \omega_{l}, {\bf q})=
\pi_{\sigma\sigma}(i \omega_{l}, {\bf q})+
\pi_{\sigma-\sigma}(i \omega_{l}, {\bf q}) $.
%
Spin and charge susceptibilities of the model (\ref{EqHam}) are
given by\cite{ComSus}
\begin{equation}\label{EqSusS}
\chi_{s}(i \omega_{l},{\bf q}) =
\frac{2\pi_{s}(i \omega_{l}, {\bf q})}
{1- \left[ U + \frac{1}{2} J^{(p)}({\bf q}) \right]
\pi_{s}(i \omega_{l}, {\bf q})} 
\end{equation}
%
and
\begin{equation}\label{EqSusC}
\chi_{c}(i \omega_{l},{\bf q}) =
\frac{2\pi_{c}(i \omega_{l}, {\bf q})}
{1 + \left[ U +  2V^{(c)}({\bf q}) \right]
\pi_{c}(i \omega_{l}, {\bf q})} ,
\end{equation}
with
$J^{(p)}({\bf q}) = \sum_{j(\ne i)} J_{ij}^{(p)}
e^{-i{\bf q} \cdot ({\bf R}_{i} - {\bf R}_{j})}$
and
$V^{(c)}({\bf q}) = \sum_{j(\ne i)} V_{ij}^{(c)}
e^{-i{\bf q} \cdot ({\bf R}_{i} - {\bf R}_{j})} $.
Irreducible polarization functions are also divided into
single-site ones,
$\tilde{\pi}_{s}(i \omega_{l})$ and 
$\tilde{\pi}_{c}(i \omega_{l})$, and multi-site ones,
$\Delta\pi_{s}(i \omega_{l}, {\bf q})$ and
$\Delta\pi_{c}(i \omega_{l}, {\bf q})$:
$\pi_{s}(i \omega_{l}, {\bf q}) =\tilde{\pi}_{s}(i \omega_{l})
+\Delta\pi_{s}(i \omega_{l}, {\bf q})$ and
$\pi_{c}(i \omega_{l}, {\bf q}) = \tilde{\pi}_{c}(i\omega_{l}) 
+\Delta\pi_{c}(i \omega_{l}, {\bf q})$.  Because the single-site
terms are  equal to those of the mapped Anderson
model,\cite{Mapping} its spin and charge susceptibilities  
are given by
\begin{equation}\label{EqSusSS}
\tilde{\chi}_{s}(i \omega_{l}) =
\frac{2\tilde{\pi}_{s}(i \omega_{l}) }
{1- U \tilde{\pi}_{s}(i \omega_{l})}
\end{equation}
and
\begin{equation}\label{EqSusSC}
\tilde{\chi}_{c}(i \omega_{l}) =
\frac{2\tilde{\pi}_{c}(i \omega_{l}) }
{1 + U \tilde{\pi}_{c}(i \omega_{l})} .
\end{equation}
The Kondo temperature is defined by
$k_{B} T_{K} \!=\! 1 / \bigl[ \tilde{\chi}_{s}(0) 
\bigr]_{T \rightarrow 0\hspace{1pt}{\rm K}}$.
When the Fermi-liquid relation\cite{Yamada,Yosida} is made use
of, it follows that 
$k_{B} T_{K} = (\tilde{\phi}_\gamma/\tilde{\phi}_s)/
\rho^*(0)$ so that $k_{B} T_{K}$ is as large as the bandwidth of
quasi-particles. In this paper, our study is restricted to the
strongly correlated regime defined by
\begin{equation}\label{EqStrongU}
k_{B}T_{K}/U \ll 1 .
\end{equation}

The Kondo temperature should be calculated by determining and
solving selfconsistently the mapped Anderson model. However,
it is treated as a phenomenological parameter in this paper.

Eqs.~(\ref{EqSusS}) and (\ref{EqSusC}) are also written in such
a way that
\begin{equation}\label{EqSusSK}
\chi_{s}(i \omega_{l},{\bf q}) =
\frac{\tilde{\chi}_{s}(i \omega_{l})  }
{1- \frac{1}{4}I_{s}(i \omega_{l}, {\bf q})
 \tilde{\chi}_{s}(i \omega_{l}) }
\end{equation}
and
\begin{equation}
\chi_{c}(i \omega_{l},{\bf q}) =
\frac{\tilde{\chi}_{c}(i \omega_{l})  }
{\displaystyle 1 + V(i \omega_{l}, {\bf q})
 \tilde{\chi}_{c}(i \omega_{l}) } , 
\end{equation}
with
\begin{equation}\label{EqIs}
I_{s}(i \omega_{l}, {\bf q}) =
J^{(p)}({\bf q})  +
\frac{ 2\Delta\pi_{s}(i \omega_{l}, {\bf q}) }
{\tilde{\pi}_{s}(i \omega_{l}) \pi_{s}(i \omega_{l}, {\bf q}) }
\end{equation}
and
\begin{equation}\label{EqIc}
V(i \omega_{l}, {\bf q}) =
V^{(c)}({\bf q})  -
\frac{ \Delta\pi_{c}(i \omega_{l}, {\bf q}) }
{2\tilde{\pi}_{c}(i \omega_{l}) \pi_{c}(i \omega_{l}, {\bf q}) }. 
\end{equation}
It follows from Eqs.~(\ref{EqSusS}) and (\ref{EqSusSS}) that
\begin{equation}\label{Eq-1/U-1}
\pi_{s}(0, {\bf q}) = \frac1{U} \left[1 + 
O \left( k_{B}T_{K}/U \right) \right]
\end{equation}
and
\begin{equation}\label{Eq-1/U-2}
\tilde{\pi}_{s}(0) = \frac1{U} \left[1 + 
O \left( k_{B}T_{K}/U \right) \right] .
\end{equation}
Here, $U \gg |J^{(p)}({\bf q})|$ is assumed. 
To leading order in $k_{B}T_{K}/U$, Eq.~(\ref{EqIs}) becomes 
\begin{equation}\label{EqExchS}
I_{s}(i \omega_{l}, {\bf q}) =
J^{(p)}({\bf q})  +
2 U^{2} \Delta\pi_{s}(i \omega_{l}, {\bf q}) .
\end{equation}
It follows from Eqs.~(\ref{EqSusC}) and (\ref{EqSusSC}) that
\begin{equation}
\left[U \!+\! V_c ({\bf q})\right] \pi_c (0,{\bf q}) =
\frac{\frac{1}{2}
\left[U\!+\!V_c ({\bf q})\right]\chi_c (0,{\bf q})}
{1 - \frac{1}{2}
\left[U\!+\!V_c ({\bf q})\right]\chi_c (0,{\bf q}) }
\end{equation}
and
$U \tilde{\pi}_c (0) = \frac{1}{2}U\tilde{\chi}_c (0) /
[1 - \frac{1}{2} U\tilde{\chi}_c (0)]$.
It is likely that
$[U \!+\! V_c ({\bf q})] |\pi_c (0,{\bf q})| \agt 1$
and 
$U |\tilde{\pi}_c (0)|\agt 1$ in actual magnets. Note also that
$\tilde{\phi}_c \ll 1$ for  $n_d\simeq 1$. Then, the second
term of Eq.~(\ref{EqIc}) can be ignored. In this paper,
Eq.~(\ref{EqIc}) is approximated by 
$V(i \omega_{l}, {\bf q}) \simeq V^{(c)}({\bf q})$. 
In $d=3$ dimensions, 
\begin{equation}\label{EqVcq}
V^{(c)} ({\bf q}) = \frac{1}{v} \sum_{\bf G} \!\biggl[
\frac{4 \pi e^2}{|{\bf q}-{\bf G}|^2} 
- \frac{1}{N} \!\sum_{\bf p (\ne 0)} \!
\frac{4 \pi e^2}{|{\bf p}-{\bf G}|^2} \biggr] ,
\end{equation}
with ${\bf G}$ wave-numbers of reciprocal lattice points,  $v$
the volume of a unit cell, and the summation over ${\bf p}$ 
restricted to the first Brillouin zone. Because 
$\lim_{{\bf q}\rightarrow{\bf G}}  V^{(c)}({\bf q})/k_BT_K = +
\infty$, the charge susceptibility vanishes for commensurate
{\bf q}:
\begin{equation}\label{EqChiCV}
\lim_{{\bf q}\rightarrow {\bf G}}
\chi_{c}(i\omega_l, {\bf q}) =0 .
\end{equation}

The susceptibility given by Eq.~(\ref{EqSusSK}) with
Eq.~(\ref{EqExchS}) is consistent with a physical picture for
Kondo lattices that local spin fluctuations at different sites
interact with each other by intersite exchange interactions.  
Then, $I_{s}(i \omega_{l}, {\bf q})$ should be called an
exchange interaction, which is responsible for magnetic
instability. The second term of Eq.~(\ref{EqExchS}) is mainly
composed of three contributions.\cite{Multi-band,OhExc}  One is
the superexchange interaction, which arises from the virtual
exchange of pair excitations across the Mott-Hubbard gap. The
phenomenological term $J^{(p)}({\bf q})$ should be determined in
such a way that the sum of this superexchange interaction and
$J^{(p)}({\bf q})$ should be equal to the superexchange
interaction of actual systems. Another is the novel exchange
interaction arising from that of pair excitations of
quasi-particles. Whether it is ferromagnetic or antiferromagnetic
depends on the dispersion relation of quasi-particles. For
example, it is ferromagnetic in the so called flat-band and
band-edge models.\cite{Multi-band}  On the other hand, it is
antiferromagnetic when the nesting of the Fermi surface is sharp
or when the chemical potential is at the band center. The other
is the mode-mode coupling term among intersite spin fluctuations.
Each of the superexchange interaction and the novel exchange
interaction can cause magnetic instability, while the
mode-mode coupling suppresses magnetic instability. In the
following part of this paper, we assume that the sum of the two
exchange interactions is antiferromagnetic and is so strong to
cause magnetic instability; the N\'{e}el temperature $T_N$ is
determined by $1/\left[\chi_s(0,{\bf Q})\right]_{T=T_N} =0$ or
$1 - \frac{1}{4}\left[I_s(0, {\bf Q})
\tilde{\chi}_s(0)\right]_{T=T_N} =0$, with ${\bf Q}$ ordering
wave-numbers.

As discussed in Introduction, local spin fluctuation are quite
different between the two temperature regions,  $T\gg T_K$ and
$T\ll T_K$.  Then, local-moment magnetism is characterized by
$T_N\gg T_K$ while itinerant-electron magnetism by $T_N\ll T_K$.
In this paper, our study is restricted to itinerant-electron
magnetism or $T_N\ll T_K$.

\subsection{Landau's free energy}\label{SecLandau}

Magnetizations appear below $T_N$. 
In this subsection, Landau's free energy in the presence
of external fields 
\begin{equation}
{\cal H}_{ext} = 
-  g\mu_{B} \sum_{i} 
 \bigl( {\bf H}_{i}  \cdot \hat{\bf s}_{i} \bigr)
 - \sum_{i} \Delta \mu_{i}  \hat{n}_{i} ,
\end{equation}
is studied. For this purpose, it is convenient to consider 
\begin{equation}\label{EqFicH}
{\cal H}_{\Omega} =
{\cal H} + E_{ext} +  {\cal H}_{L} ,
\end{equation}
instead of ${\cal H}+{\cal H}_{ext}$, with 
\begin{eqnarray}
E_{ext} &=& -  \frac{1}{2}g \mu_{B} \sum_{i} 
\left( {\bf H}_{i} \cdot {\bf m}_{i} \right)
 - \sum_{i} \Delta \mu_{i} n_{i} 
\end{eqnarray}
including no operator and
\begin{equation}\label{EqLSC}
{\cal H}_{L} = 
- \sum_{i \nu} 
\lambda_{i\nu}^{(s)} \left( 2 \hat{s}_{i\nu} \!-\! m_{i\nu} \right)
 - \sum_{i} \lambda_{i}^{(c)}  
\left(\hat{n}_{i} \!-\! n_{i} \right) 
\end{equation}
including Lagrange's multipliers,
$\lambda_{i\nu}^{(s)}$ and $\lambda_{i}^{(c)}$.
The thermodynamic potential $\Omega$ for ${\cal H}_{\Omega}$
is given by 
\begin{equation}\label{EqOmegaDef}
e^{-\Omega/k_{\rm B} T}  \equiv
{\rm Tr}~\exp \biggl[ - \frac{1}{k_{B}T} \Bigl( 
{\cal H}_{\Omega} - \mu \sum_{i} \hat{n}_{i} \Bigr)  
\biggr] ,
\end{equation}
with $\mu$ the chemical potential. Lagrange's multipliers
should satisfy
\begin{equation}\label{EqDLM}
\frac{\partial \Omega}{\partial \lambda_{i\nu}^{(s)} } = 
m_{i\nu} - 2 \langle \hat{s}_{i\nu} \rangle_{\Omega} =0 
\end{equation}
and
\begin{equation}\label{EqDLC}
\frac{\partial \Omega}{\partial \lambda_{i}^{(c)} }  = 
n_{i} - \langle \hat{n}_{i} \rangle_{\Omega}  =0 ,
\end{equation}
with $\langle \cdots \rangle_{\Omega}$ standing for thermal
averages for ${\cal H}_{\Omega}$. Then, magnetizations
and charge densities are given by sets of ${\bf m}_i$ and
$n_i$, $\{{\bf m}\}$ and $\{n\}$. It follows that
\begin{equation}\label{EqDM}
\frac{ \partial \Omega}{\partial m_{i \nu}}
=\lambda_{i\nu}^{(s)}\bigl(\{{\bf m}\},\{n\}\bigr) -
\frac{1}{2}g\mu_{B} H_{i\nu}
\end{equation}
and
\begin{equation}\label{EqDC}
\frac{ \partial \Omega}{\partial n_{i}}
= \lambda_{i}^{(c)}\bigl(\{{\bf m}\},\{n\}\bigr) -  \Delta \mu_{i}
.
\end{equation}
Here, it is not necessary to consider the dependence of $\Omega$
on $\{{\bf m}\}$ and $\{n\}$ through 
$\lambda_{i\nu}^{(s)}$ and $\lambda_{i}^{(c)}$ because of
Eqs.~(\ref{EqDLM}) and (\ref{EqDLC}).
When $\Omega$ is minimized as a function of $\{{\bf m}\}$ and
$\{n\}$, it follows  from Eqs.~(\ref{EqDM}) and (\ref{EqDC})
that $\lambda_{i\nu}^{(s)} = \frac{1}{2}g\mu_{B} H_{i\nu}$ and
$\lambda_{i}^{(c)}=\Delta \mu_{i}$, so that ${\cal H}_{\Omega}$
becomes ${\cal H}+{\cal H}_{ext}$. The set of
Eqs.~(\ref{EqDLM}) and (\ref{EqDLC}) is nothing but the
selfconsistency condition to determine magnetizations and
charge densities for ${\cal H}+{\cal H}_{ext}$, and
$\Omega$ defined this way is Landau's free energy for
${\cal H}+{\cal H}_{ext}$. 


The normal Hartree term is given by
$\lambda_{0}=-\bigl(\frac{1}{2}Un_d+\sum_j V_{ij}^{(c)}n_d\bigr)$,
with $n_d$ the average electron density defined by
Eq.~(\ref{EqND}).\cite{ComNeutraility}
All of its effects are included in the {\it unperturbed} state
considered in Sec.~\ref{SecKondoTh}. Anomalous Hartree and Fock
terms, which are due to spin or charge polarizations, 
are given by
\begin{equation}\label{EqLSR}
\overline{\lambda}_{i\nu}^{(s)} \equiv
\lambda_{i\nu}^{(s)} + \biggl( \frac{1}{2} U m_{i\nu} 
+ \frac{1}{4} \sum_{j(\ne i)} J_{ij}^{(p)} m_{j\nu} \biggr) 
\end{equation}
for spin channels, and
\begin{equation}\label{EqLCR}
\overline{\lambda}_{i}^{(c)} \equiv
\lambda_{i}^{(c)} - \biggl( \frac{1}{2} U \Delta n_{i}
+ \sum_{j(\ne i)} V_{ij}^{(c)} \Delta n_{j} \biggr) ,
\end{equation}
for the charge channel, with
\begin{equation}\label{EqDN}
\Delta n_{i} = n_{i} - n_{d} . 
\end{equation}
Here, Lagrange's multipliers, $\lambda_{i\nu}^{(s)}$ and
$\lambda_{i}^{(c)}$, are included, and thermal averages 
$\langle \hat{s}_{i\nu} \rangle_{\Omega}$ and 
$\langle \hat{n}_{i} \rangle_{\Omega}$ are replaced by
$m_{i\nu}/2$ and $n_i$ because Eqs.~(\ref{EqDLM}) and
(\ref{EqDLC}) are eventually satisfied.  When
$\overline{\lambda}_{i\nu}^{(s)}$ and
$\overline{\lambda}_{i}^{(c)}$ are perturbatively considered
starting from the Fermi-liquid state considered in
Sec.~\ref{SecKondoTh}, it follows from the selfconsistency
condition of Eqs.~(\ref{EqDLM}) and (\ref{EqDLC}) that
\begin{eqnarray}\label{EqmQL-R}
m_{i\nu} &=& 2 \sum_{j} 
\pi_{\nu^{2}}(i,j)
\overline{\lambda}_{j\nu}^{(s)}
+ 2 \sum_{j_1 j_2} \pi_{\nu^{2}c}(i,j_1.j_2)
\overline{\lambda}_{j_1\nu}^{(s)}
\overline{\lambda}_{j_2}^{(c)}
\nonumber \\ &&
+   2 \sum_{j_1 j_2} \pi_{\nu c\nu}(i,j_1,j_2)
\overline{\lambda}_{j_1}^{(c)}
\overline{\lambda}_{j_2\nu}^{(s)} 
\nonumber \\
& & + 2\sum_{j_{1}\nu_{1}}\sum_{j_{2}\nu_{2}}\sum_{j_{3}\nu_{3}}
\pi_{\nu\nu_{1}\nu_{2}\nu_{3}}(i,j_{1},j_{2},j_{3})
\nonumber \\ && \hspace*{2cm} \times
\overline{\lambda}_{j_{1}\nu_{1}}^{(s)}
\overline{\lambda}_{j_{2}\nu_{2}}^{(s)}
\overline{\lambda}_{j_{3}\nu_{3}}^{(s)} + \cdots
\end{eqnarray}
and
\begin{eqnarray}\label{EqnQL-R}
\Delta n_{i} &=&  2 \sum_{j} 
\pi_{c^{2}}(i,j) \overline{\lambda}_{j}^{(c)}
\nonumber \\ && \hspace*{-5pt}
+ 2 \sum_{j_{1}j_{2}} \sum_{\nu_{1}}
\pi_{c\nu_{1}^{2}}(i,j_{1},j_{2})
\overline{\lambda}_{j_{1}\nu_{1}}^{(s)}
\overline{\lambda}_{j_{2}\nu_{1}}^{(s)}
 + \cdots ,
\end{eqnarray}
with $\pi_{\nu^{2}}(i,j)$ static two-point polarization
functions in spin channels, $\pi_{c^{2}}(i,j)$ those in
charge channels,
$\pi_{\nu^{2}c}(i_{1},i_{2},i_{3})$,
$\pi_{\nu c\nu}(i_{1},i_{2},i_{3})$ and
$\pi_{c\nu^{2}}(i_{1},i_{2},i_{3})$ static three-point ones
with two spin vertexes and a charge vertex, and 
$\pi_{\nu_{1}\nu_{2}\nu_{3}\nu_{4}}(i_{1},i_{2},i_{3},i_{4})$
static four-point ones with four spin vertexes.

Note that
$\pi_{s}(0,{\bf q}) = \sum_{j} \pi_{\nu^{2}}(i,j)
e^{-i {\bf q} \cdot ({\bf R}_{i}-{\bf R}_{j})}$
for $\nu=x$, $y$ and $z$, and 
$\pi_{c}(0,{\bf q}) = \sum_{j} \pi_{c^{2}}(i,j)
e^{-i {\bf q} \cdot ({\bf R}_{i}-{\bf R}_{j})} $.
It follows from Eqs.~(\ref{EqmQL-R}) and (\ref{EqnQL-R}) that
in the wave-number representation 
\begin{eqnarray}\label{EqmQL} 
\lambda_{{\bf q}\nu}^{(s)} &=& 
\frac{m_{{\bf q}\nu}}{\chi_{s}(0,{\bf q})} 
- \frac{2}{\pi_{s}(0,{\bf q}) }
\sum_{{\bf q}_{1}{\bf q}_{2}}
\pi_{\nu^{2}c}({\bf q},{\bf q}_{1},{\bf q}_{2})
\overline{\lambda}_{{\bf q}_{1}\nu}^{(s)} 
\overline{\lambda}^{(c)}_{{\bf q}_{2}}
\nonumber \\ && 
- \frac{1}{\pi_{s}(0,{\bf q}) }
\sum_{\nu_{1}\nu_{2}\nu_{3}}
\sum_{{\bf q}_{1}{\bf q}_{2}{\bf q}_{3}}
\pi_{\nu\nu_1\nu_2\nu_3}
({\bf q},{\bf q}_{1},{\bf q}_{2},{\bf q}_{3})
\nonumber \\ && \hspace*{2cm} \times
\overline{\lambda}_{{\bf q}_{1}\nu_{1}}^{(s)}
\overline{\lambda}_{{\bf q}_{2}\nu_{2}}^{(s)}
\overline{\lambda}_{{\bf q}_{3}\nu_{3}}^{(s)} 
+ \cdots
\end{eqnarray}
and 
\begin{eqnarray}\label{EqnQL} 
\lambda_{{\bf q}}^{(c)} &=& 
\frac{\Delta n_{\bf q}}{\chi_{c}(0,{\bf q})} 
- \frac{1}{\pi_{c}(0,{\bf q}) }
\sum_{{\bf q}_{1}{\bf q}_{2}} 
\sum_{\nu_{1}}
\pi_{\nu_{1}^{2}c}({\bf q},{\bf q}_{1},{\bf q}_{2})
\nonumber \\ && \hspace*{3cm} \times 
\overline{\lambda}_{{\bf q}_{1}\nu_{1}}^{(s)} 
\overline{\lambda}^{(s)}_{{\bf q}_{2}\nu_{1}}
+ \cdots ,
\end{eqnarray}
with  
\begin{equation}
m_{{\bf q}\nu}=\frac{1}{N}\sum_{i}
m_{i\nu}  e^{-i{\bf q}\cdot {\bf R}_{i}},
\end{equation}
\begin{equation}
\Delta n_{\bf q}=\frac{1}{N}\sum_{i} \Delta n_{i}
e^{-i{\bf q}\cdot {\bf R}_{i}},
\end{equation}
\begin{equation}
\lambda_{{\bf q}\nu}^{(s)}=\frac{1}{N}\sum_{i}  
\lambda_{i\nu}^{(s)} e^{-i{\bf q}\cdot {\bf R}_{i}},
\end{equation}
\begin{equation}
\lambda^{(c)}_{\bf q}=\frac{1}{N}\sum_{i} 
\lambda_{i}^{(c)} e^{-i{\bf q}\cdot {\bf R}_{i}},
\end{equation}
\begin{eqnarray}
\overline{\lambda}_{{\bf q}\nu}^{(s)} &=&
\frac{1}{N} \sum_{i}
\overline{\lambda}_{i\nu}^{(s)} 
e^{-i{\bf q}\cdot {\bf R}_{i}}
\nonumber \\
&=&
\lambda_{{\bf q}\nu}^{(s)}+ \frac{1}{2} U m_{{\bf q}\nu}
+ \frac{1}{4} J^{(p)} ({\bf q}) m_{{\bf q}\nu} ,
\end{eqnarray}
\begin{eqnarray}
\overline{\lambda}^{(c)}_{\bf q} &=&
\frac{1}{N} \sum_{i}
\overline{\lambda}_{i}^{(c)} 
e^{-i{\bf q}\cdot {\bf R}_{i}}
\nonumber \\
&=&
 \lambda^{(c)}_{\bf q}- \frac{1}{2} U \Delta n_{\bf q}
-  V^{(c)} ({\bf q}) \Delta n_{\bf q} ,
\end{eqnarray}
\begin{eqnarray}\label{EqPi3}
\pi_{\nu_{1}\nu_{2}\nu_{3}}
({\bf q}_{1},{\bf q}_{2},{\bf q}_{3})  
&=& \frac{1}{N}
\sum_{i_{1}i_{2}i_{3}} \pi_{\nu_{1}\nu_{2}\nu_{3}}
(i_{1},i_{2},i_{3})
\nonumber \\ && \hspace{-1cm} \times 
e^{-i [({\bf q}_{1} \cdot {\bf R}_{{i}_{1}})
+ ({\bf q}_{2} \cdot {\bf R}_{{i}_{2}})
+ ({\bf q}_{3} \cdot {\bf R}_{{i}_{3}})] } ,
\end{eqnarray}
and
\begin{eqnarray}\label{EqPi4}
\pi_{\nu_{1}\nu_{2}\nu_{3}\nu_{4}}
({\bf q}_{1},{\bf q}_{2},{\bf q}_{3},{\bf q}_{4})  
&=& \frac{1}{N} \!\!
\sum_{i_{1}i_{2}i_{3}i_{4}} \!\!\!
\pi_{\nu_{1}\nu_{2}\nu_{3}\nu_{4}}(i_{1},i_{2},i_{3},i_{4})
\nonumber \\ && \hspace*{-3cm} \times
e^{-i[ ({\bf q}_{1} \cdot {\bf R}_{{i}_{1}})
+ ({\bf q}_{2} \cdot {\bf R}_{{i}_{2}})
+ ({\bf q}_{3} \cdot {\bf R}_{{i}_{3}})
+ ({\bf q}_{4} \cdot {\bf R}_{{i}_{4}}) ]}  .
\end{eqnarray}
Because of the translational symmetry, 
\begin{equation}
\pi_{\nu_{1}\nu_{2}\nu_{3}}
({\bf q}_{1},{\bf q}_{2},{\bf q}_{3})
\propto\delta_{{\bf q}_{1}+{\bf q}_{2}+{\bf q}_{3}} 
\end{equation}
and
\begin{equation}
\pi_{\nu_{1}\nu_{2}\nu_{3}\nu_{4}}
({\bf q}_{1},{\bf q}_{2},{\bf q}_{3},{\bf q}_{4})\propto
\delta_{{\bf q}_{1}+{\bf q}_{2}+{\bf q}_{3}+{\bf q}_{4}},
\end{equation}
with
\begin{equation}
\delta_{{\bf q}} =\left\{
\begin{array}{ll}
1, & \ {\bf q} = {\bf G} \vspace{0.2cm} \\
0, & \ {\bf q} \ne {\bf G}
\end{array} . \right.
\end{equation}

Eqs.~(\ref{EqmQL}) and (\ref{EqnQL}) can be solved in an iterative
manner. When $|m_{{\bf q}\nu}| \ll 1$ and 
$|\Delta n_{{\bf q}}| \ll 1$,  Eqs.~(\ref{EqmQL}) and
(\ref{EqnQL}) give trivial  relations, 
$\lambda_{{\bf q}\nu}^{(s)} = 
m_{{\bf q}\nu}/\chi_{s}(0,{\bf q})$
and
$\lambda_{{\bf q}}^{(c)} = 
\Delta n_{{\bf q}}/\chi_{c}(0,{\bf q})$.  
When these approximate relations are made use of, it follows that
\begin{eqnarray}\label{EqL1}
\overline{\lambda}_{{\bf q}\nu}^{(s)} 
&=& \frac{m_{{\bf q}\nu}}{2\pi_{s}(0,{\bf q})} 
\simeq \frac{1}{2}U m_{{\bf q}\nu}
\end{eqnarray}
and
\begin{eqnarray}
\overline{\lambda}^{(c)}_{\bf q} 
&=&
\frac{\Delta n_{\bf q} }{2 \pi_{c}(0,{\bf q}) } .
\end{eqnarray}
In Eq.~(\ref{EqL1}), Eq.~(\ref{Eq-1/U-1}) is also made use of. When
these  are used in Eqs.~(\ref{EqmQL}) and (\ref{EqnQL}), it follows 
that
\begin{eqnarray}\label{EqLQm}
\lambda_{{\bf q}\nu}^{(s)}  
&=&
\frac{m_{{\bf q}\nu}}{\chi_{s}(0, {\bf q})} 
- 2  \sum_{{\bf q}_{1}{\bf q}_{2}}
g({\bf q},{\bf q}_{1},{\bf q}_{2})
m_{{\bf q}_{1}\nu} 
\Delta n_{{\bf q}_{2}}
%
\nonumber \\
& &  
+ \sum_{\nu_{1}\nu_{2}\nu_{3}}
\sum_{{\bf q}_{1}{\bf q}_{2}{\bf q}_{3}}
 b_{\nu\nu_{1}\nu_{2}\nu_{3}}
 ({\bf q},{\bf q}_{1},{\bf q}_{2},{\bf q}_{3})
\nonumber \\ && \hspace*{1cm} \times 
m_{{\bf q}_{1}\nu_{1}}
m_{{\bf q}_{2}\nu_{2}}
m_{{\bf q}_{3}\nu_{3}}
+ \cdots
\end{eqnarray}
and
\begin{eqnarray}\label{EqLQn}
\lambda_{{\bf q}}^{(c)}  &=&
\frac{\Delta n_{\bf q}}{\chi_{c}(0, {\bf q})} 
- \sum_{\nu_{1}{\bf q}_{1}{\bf q}_{2}}
g({\bf q},{\bf q}_{1},{\bf q}_{2})
m_{{\bf q}_{1}\nu_{1}} m_{{\bf q}_{2}\nu_{1}} + \cdots, 
\nonumber \\ && 
\end{eqnarray}
with
\begin{equation}
g({\bf q}_{1},{\bf q}_{2},{\bf q}_{3}) =
\frac{U^{2} \pi_{x^{2}c}({\bf q}_{1},{\bf q}_{2},{\bf q}_{3})}
{2^{2}\pi_{c}(0,{\bf q}_{1})} 
\end{equation}
and
\begin{equation}\label{EqBnu}
b_{\nu_{1}\nu_{2}\nu_{3}\nu_{4}}
({\bf q}_{1},{\bf q}_{2},{\bf q}_{3},{\bf q}_{4}) =
- \frac{U^{4}}{2^{3}} 
\pi_{\nu_{1}\nu_{2}\nu_{3}\nu_{4}}
 ({\bf q}_{1},{\bf q}_{2},{\bf q}_{3},{\bf q}_{4}) .
\end{equation}

Eqs.(\ref{EqDM}) and (\ref{EqDC}) can also be written in the
momentum representation, and Lagrange's multipliers are
given by Eqs.~(\ref{EqLQm}) and (\ref{EqLQn}). By integrating
them, we obtain
\begin{equation}
\Omega = \Omega_{0} + \Omega_{ext}+  \Omega_{s^2} 
+ \Omega_{c^2} + \Omega_{s^2c} + \Omega_{s^4}  + \cdots,
\end{equation}
with $\Omega_{0}$ the term including 
no spin nor charge polarizations, 
\begin{equation}
\Omega_{ext} = -  N \biggl[ 
\frac{1}{2}g \mu_{B} 
\sum_{\bf q} 
\bigl({\bf H}_{\bf q} \!\cdot\! {\bf m}_{-{\bf q}} \bigr)
 +  \sum_{\bf q} \Delta \mu_{\bf q} \Delta n_{-{\bf q}}  
\biggr],
\end{equation}
\begin{equation}
\Omega_{s^2} = N \sum_{\bf q}
\frac{|{\bf m}_{\bf q}|^{2}}{2 \chi_{s}(0,{\bf q})},
\end{equation}
\begin{equation}
 \Omega_{c^2} =
N  \sum_{\bf q}
\frac{|\Delta n_{\bf q}|^{2}}{2 \chi_{c}(0,{\bf q})} ,
\end{equation}
\begin{eqnarray}\label{EqOS3}
 \Omega_{s^2c} &=& -  N \!\!
\sum_{{\bf q}_{1}{\bf q}_{2}{\bf q}_{3}} \!
g({\bf q}_{1},{\bf q}_{2},{\bf q}_{3})
%
({\bf m}_{{\bf q}_{1}} \!\cdot\! {\bf m}_{{\bf q}_{2}} )
\Delta n _{{\bf q}_{3}} ,
\end{eqnarray}
and
\begin{eqnarray}\label{EqOS4}
\Omega_{s^4} &=& \frac{N}{4}
\sum_{{\bf q}_{1}{\bf q}_{2}{\bf q}_{3}{\bf q}_{4}} 
b_{\nu^{4}}({\bf q}_{1},{\bf q}_{2},{\bf q}_{3},{\bf q}_{4})
\nonumber \\ && \hspace{-0.5cm} \times 
\Bigl[  
({\bf m}_{{\bf q}_{1}} \!\cdot {\bf m}_{{\bf q}_{2}} )
({\bf m}_{{\bf q}_{3}} \!\cdot {\bf m}_{{\bf q}_{4}})
%
- ({\bf m}_{{\bf q}_{1}} \!\cdot {\bf m}_{{\bf q}_{3}} )
({\bf m}_{{\bf q}_{2}} \!\cdot {\bf m}_{{\bf q}_{4}})
\nonumber \\
&& \hspace*{1cm} 
+ ({\bf m}_{{\bf q}_{1}} \!\cdot {\bf m}_{{\bf q}_{4}} )
({\bf m}_{{\bf q}_{2}} \!\cdot {\bf m}_{{\bf q}_{3}})
\Bigr]  .
\end{eqnarray}
In Eq.~(\ref{EqOS4}), the following relations of
\begin{eqnarray}
b_{\nu^4}({\bf q}_{1},{\bf q}_{2},{\bf q}_{3},{\bf q}_{4})  
&=&
b_{\nu_1^2\nu_2^2}
({\bf q}_{1},{\bf q}_{2},{\bf q}_{3},{\bf q}_{4})
\nonumber \\ &=&
- \Bigl[ b_{\nu_{1}\nu_{3}\nu_{1}\nu_{3}}
({\bf q}_{1},{\bf q}_{2},{\bf q}_{3},{\bf q}_{4})
\Bigr]_{\nu_{1} \ne \nu_{3}} \phantom{\Biggr]}
\nonumber \\ && 
\end{eqnarray}
are made use of; $b_{\nu_{1}\nu_{2}\nu_{3}\nu_{4}} 
({\bf q}_{1},{\bf q}_{2},{\bf q}_{3},{\bf q}_{4})=0$
for other combinations of $\nu_i$'s.

When ordering wave-numbers are incommensurate,
there can be several equivalent ones to them. Denote the
$i$\hspace{1pt}th pair by $\pm{\bf Q}_{i}$, and assume that
not only $\pm{\bf Q}_{i}$ but also $\pm 2 {\bf Q}_{i}$ and $\pm
4 {\bf Q}_{i}$ are incommensurate:  
$\pm 2{\bf Q}_{i} \ne {\bf G}$ or  
$\pm 4{\bf Q}_{i} \ne {\bf G}$. When SDW with 
$\pm {\bf Q}_{i}$ are stabilized,  CDW with $\pm 2{\bf Q}_{i}$ 
appear because of the coupling term  between SDW and CDW
given by Eq.~(\ref{EqOS3}) or
\begin{eqnarray}\label{EqS2-C}
\Omega_{s^2c} &=&
-  N \sum_i g({\bf Q}_i,{\bf Q}_i,- 2{\bf Q}_i)
\nonumber \\ && \times \Bigl[
\left({\bf m}_{{\bf Q}_i}\right)^2
\Delta n_{-2{\bf Q}_i}  +
\left({\bf m}_{-{\bf Q}_i}\right)^2
\Delta n_{2{\bf Q}_i} \Bigr] .
\end{eqnarray}
Here and in the following part, the summation over ${\bf Q}_i$
is made over pairs or it is made in such a way that  one
of $\pm{\bf Q}_i$ is only considered for each $i$.  When only SDW
or helical structures with ${\bf Q}_{i}$ and CDW with
$2{\bf Q}_{i}$ are considered, 
\begin{eqnarray}
\Omega_{ext} &=& -N \sum_{i} \biggl\{
\frac{1}{2}g \mu_{B} 
\Bigl[ \bigl({\bf H}_{{\bf Q}_{i}} \!\cdot\! 
{\bf m}_{-{\bf Q}_{i}}\bigr)
+\bigl({\bf H}_{-{\bf Q}_{i}} \!\cdot\!
{\bf m}_{{\bf Q}_{i}}\bigr) \Bigr] 
\nonumber \\
&& + \Bigl[
\Delta \mu_{2{\bf Q}_{i}} \Delta n_{-2{\bf Q}_{i}} 
+ \Delta \mu_{-2{\bf Q}_{i}} \Delta n_{2{\bf Q}_{i}}
\Bigr] \biggr\} ,
\end{eqnarray}
\begin{equation}\label{EqOmegaS2}
\Omega_{s^2} = N \sum_{i} 
\frac{|{\bf m}_{{\bf Q}_{i}}|^2} {\chi_{s}(0,{\bf Q}_i)},
\end{equation}
\begin{equation}
\Omega_{c^2} = N \sum_{i} 
\frac{|\Delta n_{2{\bf Q}_{i}}|^2}
{\chi_{c}(0,2{\bf Q}_i)} ,
\end{equation}
and
\begin{eqnarray}\label{EqOmegaS4}
\Delta\Omega_{s^4} &=&
\frac{N}{2} \sum_{i} \Bigl\{ 
B_{1} \left[ 2 |{\bf m}_{{\bf Q}_{i}} |^{4} -  
|({\bf m}_{{\bf Q}_{i}} \!\cdot {\bf m}_{{\bf Q}_{i}} )|^{2} 
\right] 
\nonumber \\ && \hspace*{1.2cm} + 
2 B_{2}
|({\bf m}_{{\bf Q}_{i}} \!\cdot {\bf m}_{{\bf Q}_{i}} )|^{2} 
\Bigr\} 
\nonumber \\ && 
+  \frac{N}{2} \sum_{i \ne j} \Bigl\{ 2B_{3}
|{\bf m}_{{\bf Q}_{i}} |^{2}  |{\bf m}_{{\bf Q}_{j}} |^{2}
\nonumber \\ && \quad  +
2B_{4} \left[
|({\bf m}_{{\bf Q}_{i}} \!\cdot {\bf m}_{{\bf Q}_{j}})|^{2}
+ |({\bf m}_{{\bf Q}_{i}} \!\cdot {\bf m}_{-{\bf Q}_{j}})|^{2}
\right]
\nonumber \\ && \quad +
B_{5} \left[
|({\bf m}_{{\bf Q}_{i}} \!\cdot {\bf m}_{{\bf Q}_{j}})|^{2}
-|({\bf m}_{{\bf Q}_{i}} \!\cdot {\bf m}_{-{\bf Q}_{j}})|^{2}
\right]\! \Bigr\} ,
\nonumber \\ && \phantom{\Bigr]} 
\end{eqnarray}
with
\begin{equation}
B_{1} = b_{\nu^{4}}({\bf Q},-{\bf Q},{\bf Q},-{\bf Q}) ,
\end{equation}
\begin{equation}
B_{2} =
 b_{\nu^{4}}({\bf Q},{\bf Q},-{\bf Q},-{\bf Q}) ,
\end{equation}
\begin{eqnarray}
B_{3} &=&
b_{\nu^{4}}({\bf Q},-{\bf Q},{\bf Q}^\prime,-{\bf Q}^\prime)
+ b_{\nu^{4}}({\bf Q},-{\bf Q},-{\bf Q}^\prime,{\bf Q}^\prime)
\nonumber \\ && 
- \frac{1}{2} 
b_{\nu^{4}}({\bf Q},{\bf Q}^\prime,-{\bf Q},-{\bf Q}^\prime)
- \frac{1}{2} 
b_{\nu^{4}}({\bf Q},-{\bf Q}^\prime,-{\bf Q},{\bf Q}^\prime),
\nonumber \\ && 
\end{eqnarray}
\begin{equation}
B_{4} = \frac{1}{2} 
b_{\nu^{4}}({\bf Q},{\bf Q}^\prime,-{\bf Q},-{\bf Q}^\prime)
+ \frac{1}{2} 
b_{\nu^{4}}({\bf Q},-{\bf Q}^\prime,-{\bf Q},{\bf Q}^\prime) ,
\end{equation}
and
\begin{equation}
B_{5} =
2 b_{\nu^{4}}({\bf Q},-{\bf Q},{\bf Q}^\prime,-{\bf Q}^\prime)
-2b_{\nu^{4}}
({\bf Q},-{\bf Q},-{\bf Q}^\prime,{\bf Q}^\prime) .
\end{equation}
Here, ${\bf Q}$ are ${\bf Q}^\prime$ 
$({\bf Q}\ne\pm{\bf Q}^\prime)$ are two different
wave-numbers among considered ${\bf Q}_i$'s, and symmetrical
relations such as
%
$b_{\nu^{4}}({\bf q}_{1},{\bf q}_{2},{\bf q}_{3},{\bf q}_{4})
=b_{\nu^{4}}({\bf q}_{4},{\bf q}_{1},{\bf q}_{2},{\bf q}_{3})=
b_{\nu^{4}}(-{\bf q}_{1},-{\bf q}_{2},-{\bf q}_{3},-{\bf q}_{4})
$
%
and so on, are made use of.

Four-point polarization functions are also divided into
single and multisite terms so that
\begin{equation}
b_{\nu^{4}}({\bf q}_{1},{\bf q}_{2},{\bf q}_{3},{\bf q}_{4})
= \tilde{b}_{\nu^{4}} + 
\Delta 
b_{\nu^{4}}({\bf q}_{1},{\bf q}_{2},{\bf q}_{3},{\bf q}_{4}) .
\end{equation}
The single-site term is approximately given by
Eq.~(\ref{EqBNu4}):
$\tilde{b}_{\nu^{4}}=2/\tilde{\chi}_s(0)=2k_BT_K$. When a diagram
corresponding to Fig.~2$(b)$ of the previous paper\cite{OhSDW-CW}
is considered and Eq.~(\ref{EqGreen}) is approximately used, the
multi-site term is calculated so that for $T \ll T_K$ 
\begin{eqnarray}\label{EqDBNu4}
&&\Delta 
b_{\nu^{4}}({\bf q}_1,{\bf q}_2,{\bf q}_3,{\bf q}_4) =
\delta_{{\bf q}_{1}+{\bf q}_{2}+{\bf q}_{3}+{\bf q}_{4}} 
\hspace{2pt} 2 \hspace{-1pt} \left(\hspace{-2pt}
\frac{\tilde{\phi}_s}{\tilde{\phi}_\gamma\tilde{\chi}_s(0)}
\hspace{-2pt}\right)^4 \hspace{-3pt} \times 
\nonumber \\ && \hspace{1.3cm} \times
k_BT_K \sum_{n}\frac{1}{N} \sum_{\bf k} 
\frac{1}{i\varepsilon_n \!-\! \xi({\bf k})}
\frac{1}{i\varepsilon_n \!-\! \xi({\bf k}\!+\!{\bf q}_1)}
\nonumber \\ && \hspace{1.3cm} \times 
\frac{1}{i\varepsilon_n \!-\! 
\xi({\bf k}\!+\!{\bf q}_1\!+\!{\bf q}_2)}
\frac{1}{i\varepsilon_n \!-\! 
\xi({\bf k}\!+\!{\bf q}_1\!+\!{\bf q}_2\!+\!{\bf q}_3)}.
\nonumber \\ && \phantom{\big]} 
\end{eqnarray}
Here, the Ward relation for the three-point vertex function in
spin channels is made use of, as it was in the previous
paper.\cite{OhSDW-CW} In general, 
\begin{equation}
|B_{5}| \ll {\rm Min}\left(B_1, B_{2},B_{3},B_{4}\right).
\end{equation}
The nesting of the Fermi surface is characterized by
$\xi({\bf k})\simeq -\xi({\bf k}+{\bf Q})$ in a wide {\bf k} region
in the vicinity of the Fermi surface. When the nesting of the
Fermi surface is sharp, $B_1$ is much lager than other $B_i$'s
so that 
\begin{equation}
B_{1}\gg{\rm Max}\left(B_{2},B_{3},B_{4}\right) \gg |B_5|.
\end{equation}

\section{incommensurate SDW and CDW}\label{MagStr}
\subsection{Sinusoidal or helical structures}

In this subsection, it is examined which are
stabilized, sinusoidal or helical structures. The second order
term in magnetizations, $\Delta \Omega_{s^2}$, does not depend on
magnetic structures. When magnetizations are small, magnetic
structures are determined by the fourth order term
$\Delta\Omega_{s^4}$.
When single ${\bf Q}$ structures are assumed, the fourth order
term is written as
\begin{eqnarray}
\Delta \Omega_{s^4} &=&
\frac{N}{2}  \Bigl\{ B_{1}
\left[  2 |{\bf m}_{\bf Q} |^{4} -  
|({\bf m}_{\bf Q} \!\cdot {\bf m}_{\bf Q} )|^{2} \right] 
\nonumber \\
&&  \hspace*{1cm} +  2 B_{2} |({\bf m}_{\bf Q} \!\cdot 
{\bf m}_{\bf Q} )|^{2} \Bigr\} .
\end{eqnarray}
When magnetizations are written as 
${\bf m}_{\bf Q} = {\bf m}_{\bf Q}^{\prime} 
+  i \hspace{2pt} {\bf m}_{\bf Q}^{\prime\prime} $,
with 
${\bf m}_{\bf Q}^{\prime}$ and ${\bf m}_{\bf Q}^{\prime\prime}$
being real, 
it follows that
\begin{equation}
({\bf m}_{\bf Q} \cdot {\bf m}_{\bf Q}) =
\left({\bf m}_{\bf Q}^{\prime} \right)^2 
- \left({\bf m}_{\bf Q}^{\prime\prime}\right)^2
+ i 2 ({\bf m}_{\bf Q}^{\prime} \cdot 
{\bf m}_{\bf Q}^{\prime\prime}) .
\end{equation}
It is easy to see that 
$|({\bf m}_{\bf Q} \cdot {\bf m}_{\bf Q})|^2$ becomes the
smallest for helical structures, which are characterized by 
$|{\bf m}_{\bf Q}^{\prime}| = |{\bf m}_{\bf Q}^{\prime\prime}|$
and
${\bf m}_{\bf Q}^{\prime} \perp {\bf m}_{\bf Q}^{\prime\prime}$,
such as
\begin{equation}\label{EqHelM}
|({\bf m}_{\bf Q} \cdot {\bf m}_{\bf Q})|^2 =0 .
\end{equation} 
On the other hand, it becomes the
largest for  sinusoidal structures, which are characterized by
${\bf m}_{\bf Q}^{\prime} \parallel {\bf m}_{\bf
Q}^{\prime\prime}$, such as 
\begin{equation}\label{EqSinM}
|({\bf m}_{\bf Q} \cdot {\bf m}_{\bf Q})|^2 =
|{\bf m}_{\bf Q}|^4.
\end{equation}
Therefore, helical structures are stabilized when
\begin{equation}\label{EqCondHelical}
B_1 - 2B_2 < 0 ,
\end{equation}
and sinusoidal structures are stabilized when
\begin{equation}\label{EqCondSDW}
B_1 - 2B_2 > 0 .
\end{equation}

When no multi-site terms are included,
$B_{1} = B_{2} = B_{3} = B_{4} = \tilde{b}_{\nu^{4}}$ and
$B_{5}=0$. When there is no sharp nesting of the Fermi surface,
therefore, it is likely that Eq.~(\ref{EqCondHelical}) is satisfied.
The novel exchange interaction can never be strongly
antiferromagnetic in this case.\cite{Multi-band,OhExc} However,
antiferromagnetic instability can still occur when the
superexchange interaction is strongly antiferromagnetic. Even
helical structures are stabilized in such itinerant-electron
antiferromagnets. When the nesting of the Fermi surface is so
sharp for incommensurate {\bf Q} that Eq.~(\ref{EqCondSDW}) might
be satisfied, sinusoidal structures are stabilized as is
expected.

\subsection{Single or multiple {\bf Q}  structures}

In this subsection, it is examined whether multiple {\bf Q}
structures are stabilized or not.  First, examine helical
structures under an assumption that Eq.~(\ref{EqCondHelical}) is
satisfied. Because
$({\bf m}_{\bf Q} \cdot {\bf m}_{\bf Q}) =0$ for helical
structures, $\Delta\Omega_{s^4}$ is reduced to
\begin{eqnarray}
\Delta\Omega_{s^4} &=&
\frac{N}{2} \sum_{i} B_{1}
|{\bf m}_{{\bf Q}_{i}} |^{4}  
%
+ \frac{N}{2} \sum_{i \ne j} \Big\{
2B_{3}
|{\bf m}_{{\bf Q}_{i}} |^{2}  |{\bf m}_{{\bf Q}_{j}} |^{2}
\nonumber \\ && 
+ 2B_{4} \left[
|({\bf m}_{{\bf Q}_{i}} \!\cdot {\bf m}_{{\bf Q}_{j}})|^{2}
+ |({\bf m}_{{\bf Q}_{i}} \!\cdot {\bf m}_{-{\bf Q}_{j}})|^{2}
\right]
\nonumber \\ && 
+ B_{5} \left[
|({\bf m}_{{\bf Q}_{i}} \!\cdot {\bf m}_{{\bf Q}_{j}})|^{2}
- |({\bf m}_{{\bf Q}_{i}} \!\cdot {\bf m}_{-{\bf Q}_{j}})|^{2}
\right] \! \Big\} .
\nonumber \\ && 
\end{eqnarray}
When Eq.~(\ref{EqCondHelical}) is satisfied, it is likely that
$B_{1} < 2B_{3}$, $B_{1} < 2B_{4}$, and $|B_{5}|$ is much smaller
than other $B_{i}$.  In such a case, multiple helical structures
are never stabilized. When helical structures appear at
temperatures much lower than $T_K$, they  must be of single {\bf
Q}.

Next, examine sinusoidal structures under an assumption that
Eq.~(\ref{EqCondSDW}) is satisfied. In this case, we can put
${\bf m}_{{\bf Q}_{i}} = e^{i \theta_{i}} \bar{\bf m}_i$ 
and
${\bf m}_{-{\bf Q}_{i}} = e^{-i \theta_{i}} \bar{\bf m}_i$,
with $\bar{\bf m}_i$ being real.  The
fourth order term is given by
\begin{eqnarray}
\Omega_{s^4} &=& 
\frac{N}{2} \sum_{i} \left( B_{1}+  2 B_{2} \right)
|\bar{\bf m}_i|^{4} 
\nonumber \\ && 
+  N \sum_{i \ne j} \Bigl[ B_{3}
|\bar{\bf m}_i|^{2} |\bar{\bf m}_j|^{2}
+ 2B_{4} (\bar{\bf m}_i \!\cdot \bar{\bf m}_j)^{2} \Bigr] .
\end{eqnarray}
When the multiplicity of ${\bf Q}$ is two or three,  
$\Omega_{s^4}$ becomes the smallest when 
\begin{equation}\label{EqMperpM}
(\bar{\bf m}_i \cdot \bar{\bf m}_j)= 0
\end{equation}
for any pair of $i\ne j$. 
When Eq.~(\ref{EqCondSDW}) is satisfied, it is likely  that
\begin{equation}
B_{1} + 2 B_{2}  > 2 B_{3}
\end{equation}
is also satisfied. Then, multiple sinusoidal structures are
stabilized and their polarization vectors are orthogonal to each
other. 
It follows that
\begin{equation}
\Omega_{s^2}+\Omega_{s^4} = N \lambda \left[
\frac{m^2}{\chi_{s}(0, {\bf Q}) } 
+ \frac{1}{2} B_\perp (\lambda)m^{4} \right] ,
\end{equation}
with
\begin{equation}
B_\perp (\lambda) = B_{1} +  2 B_{2}
+ 2 (\lambda \!-\! 1) B_{3}.
\end{equation}
Here,  $|{\bf m}_i|$ is simply denoted by $m$ because
all of $|{\bf m}_i|$ are the same as each other, and $\lambda$ is
the multiplicity of ordering wave-numbers, $\lambda=1, 2$ or 3.
Below $T<T_{N}$, 
\begin{equation}
m^2 = - \frac{1}
{B_\perp (\lambda) \chi_{s}(0, {\bf Q}) } 
\end{equation}
and
\begin{equation}\label{EqOmega3}
\Omega_{s^2} + \Omega_{s^4} =
- \frac{\lambda N}
{2 B_\perp (\lambda) |\chi_{s}(0, {\bf Q}) |^{2}} .
\end{equation}
Because Eq.~(\ref{EqOmega3}) is an increasing function of 
$\lambda$, a symmetry broken ordered state is never stabilized
as long as $\lambda \le 3$; if there are three equivalent {\bf
Q}'s, for example, single or double {\bf Q} structures are
never stabilized.

When the multiplicity of ${\bf Q}$ is four or larger than four 
$(\lambda \ge 4)$, it is impossible that Eq.~(\ref{EqMperpM}) is
satisfied for any pair of ${\bf Q}_i$ and ${\bf Q}_j$. Only two
possible magnetic structures are examined in this paper; triple or
quartet ${\bf Q}$ structures.  Triple structures are
symmetry broken states, where only three ${\bf Q}$'s among
$\lambda$ wave-numbers are ordered and their magnetizations are
orthogonal to each other. Their free energy is given by
Eq.~(\ref{EqOmega3}) with $\lambda=3$. In quartet structures,
four ${\bf Q}$ components are ordered and each pair of
magnetizations make the same angle, $\cos^{-1}(-1/3) = 121.63
\cdots$~degree;  for example, 
$\bar{\bf m}_1 = m\left( 0,0,1\right)$, 
$\bar{\bf m}_2 = m\left( 0,\sqrt{8}/3,-1/3\right)$,
$\bar{\bf m}_3 = m\left( \sqrt{6}/3,-\sqrt{2}/3,-1/3\right)$,
and
$\bar{\bf m}_4 = m\left( -\sqrt{6}/3,-\sqrt{2}/3,-1/3\right)$.
Then, it follows that 
\begin{equation}
\Omega_{s^2} + \Omega_{s^4} = 4N \left[
\frac{m^2}{\chi_{s}(0, {\bf Q}_{i})}
+ \frac{1}{2} B_Q \hspace{1pt}m^{4} \right]
\end{equation}
with
\begin{equation}
B_{Q} = B_{1} + 2B_{2} + 6B_{3}+ \frac{4}{3}B_{4} .
\end{equation}
When the free energy is minimized, 
\begin{equation}
\Omega_{s^2} + \Omega_{s^4}  = - \frac{4N}
{2B_Q |\chi_{s}(0, {\bf Q}) |^{2}}
\end{equation}
below $T_{N}$. When $4/B_Q  >3/B_\perp(3) $ or
\begin{equation}\label{EqT-Q}
B_{1} > 4B_{4} - 2 B_{2} + 2 B_{3} ,
\end{equation}
is satisfied, quartet {\bf Q} structures are
stabilized. Otherwise, triple {\bf Q} structures are
stabilized. One can conclude that quartet {\bf Q} structures can
be stabilized only when the nesting of the Fermi surface is
extraordinarily sharp.

\subsection{Coexistence of SDW and CDW}
\label{SecSCDW}

The coupling between helical structures and CDW vanishes
according to Eqs.~(\ref{EqS2-C}) and (\ref{EqHelM}), and  
that between SDW and CDW exists
according to Eqs.~(\ref{EqS2-C}) and (\ref{EqSinM}).
Once SDW with $\pm {\bf Q}$ appear, therefore, CDW with $\pm
2{\bf Q}$ appear. The amplitude of CDW is given by
\begin{equation}
\Delta n_{-2{\bf Q}} = g({\bf Q},{\bf Q},- 2{\bf Q})
 \chi_{c}(0,2{\bf Q}) 
{\bf m}_{\bf Q}^{2} .
\end{equation}
When multi-site terms are ignored, 
$g({\bf Q},{\bf Q},- 2{\bf Q})$ is approximately given by
Eq.~(\ref{Eq-g}):
\begin{equation}
g({\bf Q},{\bf Q},- 2{\bf Q}) \simeq 
\left\{\begin{array}{cc}
\Delta /\pi, &  \  n_d <1
\vspace{0.2cm}\\ 
- \Delta /\pi, & \  n_d >1
\end{array} \right. ,
\end{equation}
with $\Delta$ the hybridization energy of the mapped Anderson
model.  According to the mapping condition,\cite{Mapping}
$1/\Delta \simeq - \mbox{Im}(1/N) 
\sum_{\bf k}G_\sigma (+i0,{\bf k})$, so that the coupling constant
is rather large. However, the charge susceptibility is small. Then,
the amplitude of CDW must be  small. In particular, it vanishes 
when $2{\bf Q}$ is commensurate because of Eq.~(\ref{EqChiCV}).

Denote polarizations as 
${\bf m}_{\bf Q} = \frac{1}{2}{\bf m} e^{i \theta_{s}} $
and 
$\Delta n_{2{\bf Q}} = \frac{1}{2}|\Delta n|e^{i \theta_{c}}$,
with ${\bf m}$ being real.
Then, the coupling term is given by
\begin{eqnarray}
\Delta \Omega_{s^2c} &=& 
-\frac{1}{2} Ng({\bf Q},{\bf Q},- 2{\bf Q})
\chi_{c}(0, {\bf Q}) 
\nonumber\\&&\quad\quad\times
|{\bf m}|^{2}|\Delta n| 
\cos (2\theta_{s}- \theta_{c}) .
\end{eqnarray}
The free energy takes its minimum when 
\begin{equation}
2\theta_{s}- \theta_{c} = 
\left\{ \begin{array}{cc}
2 l \pi, & n_d <1 \vspace{0.2cm} \\
(2l+1) \pi, & n_d > 1 
\end{array} \right. ,
\end{equation}
with $l$ being integers. Then, it follows that
\begin{equation}\label{EqSDW-A}
{\bf m}_{i} = {\bf m}
\cos ( {\bf Q}\cdot {\bf R}_{i} + \theta_{s})
\end{equation}
and
\begin{eqnarray}\label{EqCDW-A}
\Delta n_{i} =\left\{
\begin{array}{cc}
+ |\Delta n|
\cos\bigl[2({\bf Q}\cdot {\bf R}_{i}+\theta_{s})\bigr] ,
& n_d <1 \vspace{0.2cm} \\
- |\Delta n|
\cos\bigl[2({\bf Q}\cdot {\bf R}_{i}+\theta_{s})\bigr] ,
& n_d >1 
\end{array} \right. 
\end{eqnarray}

The quenching of magnetic moments by local quantum spin
fluctuations, which is one of the most essential effects in Kondo
lattices, sensitively depends on local number of electrons. 
According to Eq.~(\ref{EqLocalTK}) in Appendix, the Kondo
temperature at the $i$\hspace{1pt}th site is approximately given
by 
\begin{equation}\label{EqTK}
\bigl[k_BT_K\bigr]_i = \frac{2\Delta}{\pi}
\bigl|1 - (n_d + \Delta n_i) \bigr|.
\end{equation}
Eqs.~(\ref{EqSDW-A}), (\ref{EqCDW-A}) and (\ref{EqTK}) show that
electron numbers $\Delta n_{i}$ are modulated in such a way that
magnetic moments are much quenched where $T_{K}$ are high. In the
less than half-filled case $(n_d<1)$, for example, doped holes go
mainly into sites where magnetic moments are small.

\section{Discussion}\label{Discussion}
\label{SecDiscussion}

Because no magnetic anisotropy is taken into account, directions
of magnetic polarizations are totally independent of those of
wave-numbers {\bf Q} in this paper. In actual magnets, absolute
directions of magnetizations are mainly determined by magnetic
anisotropy. However, the relative angles of magnetizations
between different {\bf Q} components must be 90$^\circ$ or at
least close to 90$^\circ$ in double or triple {\bf Q} structures
even if magnetic anisotropy is taken into account; the magnetic
part of the free energy is lower when the relative angles  are
closer to 90$^\circ$.

In Kondo lattices, the coupling between SDW and CDW is strong
and the phases of SDW and CDW are never independent of
each other.  Then, even nonmagnetic impurities have a large
pinning effect on coexisting SDW and CDW.

Incommensurate SDW were observed in many metallic magnets, some
of which are of high enough symmetry such as cubic
CeAl$_2$,\cite{Barbara,Steglich} cubic Cr,\cite{Arrot,Werner} 
cuprate-oxide high-temperature superconductors,\cite{Tranquada}
which are approximately regarded as orthorhombic lattices, and so
on. In this paper, Landau's free energy is obtained up to the
fourth order in magnetizations. Because  saturated magnetic
moments are large in CeAl$_2$ and Cr, higher than the fourth
order terms are required to discuss their physical properties at
$T \ll T_N$. However, it is still interesting to examine whether
or not a multiple {\bf Q} structure is actually stabilized in
CeAl$_2$ because magnetic moments are small just below $T_N$.  It
should be mentioned that a triple {\bf Q} structure was
actually proposed for CeAl$_2$.\cite{Shapiro,Mac} The transition
at $T_N \simeq 38^\circ$C in Cr is of first order, and  it is not
certain if magnetizations are small enough even just below $T_N$.
Because a multiple {\bf Q} structure was also
suggested,\cite{Weiss} it is interesting to reexamine SDW in
cubic Cr. It has been claimed that the so called stripe order
must be stabilized in the cuprate oxides.\cite{Tranquada} Because
magnetizations are small, the theory of this paper is applicable
to magnetic states in the cuprate oxides. Within the theoretical
framework of this paper, the stripe order must be nothing but
the coexisting state of incommensurate SDW and CDW. It is
interesting  which is actually responsible for two equivalent
satellites in neutron diffraction, two equivalent magnetic
domains whose volumes are accidentally almost the same as each
other or a double {\bf Q} structure of SDW and CDW.

The spectral weight of Gutzwiller's quasi-particle band is
small in strongly correlated electron systems in the vicinity
of the Mott transition. A large part of the spectral weight
exists in the lower and upper Hubbard bands, which are far
from the chemical potential. The formation of SDW and CDW
causes not only gaps in  quasi-particle spectra but also
pseudogaps in the lower and upper Hubbard bands.  Their
pseudogaps are as large as $U|{\bf m}({\bf Q})|$, although 
gaps in quasi-particle spectra are of the order of $k_BT_N$. It
is interesting to observe changes in the density of states
caused by the formation of SDW and CDW not only in the
vicinity of the chemical potential but also far from the
chemical potential.

\section{Conclusion}\label{Conclusion}
\label{SecConclusion}

Magnetic and charge structures in strongly correlated electron
liquids are studied within the theoretical framework of Kondo
lattices. When there is no sharp nesting in the Fermi surface
and the superexchange interaction is strongly
antiferromagnetic, helical structures are stabilized even in
itinerant-electron magnets. When the nesting of the Fermi
surface is sharp enough, sinusoidal structures are
stabilized.  When incommensurate ordering wave-numbers are  
multiple, in particular, multiple sinusoidal structures are
stabilized. Their magnetic polarizations are orthogonal to
each other in double or triple sinusoidal structures, when
magnetic anisotropy is small enough.

Because the quenching of magnetic moments by local quantum spin
fluctuations sensitively depends on local numbers of electrons
in Kondo lattices, incommensurate CDW is inevitably driven by
incommensurate SDW. In hole or electron doped systems at the
vicinity of the Mott transition, therefore, the phase of CDW is
adjusted in such a way that the density of doped carriers, holes
or electrons, is large at sites where magnetic moments are
small.

\begin{acknowledgments}
The author is thankful to K. Fujiwara for showing him  the
references.\cite{Barbara,Steglich,Shapiro,Mac} This work was
supported by a Grant-in-Aid for Scientific Research (C)
No.~13640342 from the Ministry of Education, Cultures, Sports,
Science and Technology of Japan.
\end{acknowledgments}

\appendix
\section{Non-mapped single-impurity Anderson model}
\label{Anderson}

Consider the Anderson model with infinitely large $U$,
constant hybridization energy $\Delta$, and conduction
bandwidth $2D$. According to a previous paper,\cite{Yamamoto} the
ground state energy of such an Anderson model is given by
\begin{equation}
E_{G} = \epsilon_{d} - \mu -\tilde{\lambda}_{c} -
\sqrt{ (k_BT_K)^2 e^{-\pi \tilde{\lambda}_{c} /\Delta} 
+ \sum_{\nu} \tilde{\lambda}_{\nu}^{2} } ,
\end{equation}
with $k_B T_{K} = \sqrt{D \Delta} \exp \left[ 
\pi( \epsilon_{d} - \mu )/2 \Delta \right]$.
Here, $\epsilon_{d}$ is the energy
level of strongly correlated electrons and $\mu$ is the
chemical potential; $\tilde{\lambda}_{c}$ and
$\tilde{\lambda}_{\nu}$ ($\nu=x, y$ and $z$) are external
fields or Lagrange multipliers, which are determined in such a
way that magnetic moments and electron numbers are 
${\bf m}=(m_x,m_y.m_z)$ and $n$.

In the absence of fields, the electron number is given by
$n_{d} = - \left[ \partial E_{G}/\partial \lambda_{c}
\right]_{\tilde{\lambda}_{c}=0,\tilde{\lambda}_{\nu}=0}
= 1 - (\pi T_{K}/2 \Delta) $, so that  
\begin{equation}\label{EqLocalTK}
k_B T_K = \frac{2 \Delta}{\pi} (1-n_d) .
\end{equation}
In the presence of fields, magnetic moments and the electron 
number are given by
\begin{equation}\label{EqAndM}
m_{\nu} = - \frac{\partial E_{G}}{\partial \tilde{\lambda}_{\nu}}
= \frac{\tilde{\lambda}_{\nu}}
{\sqrt{T_{K}^{2}e^{-\pi \tilde{\lambda}_{c} /\Delta} 
 + \sum_{\nu} \tilde{\lambda}_{\nu}^{2} }} 
\end{equation}
and
\begin{equation}\label{EqAndN}
n = - \frac{\partial E_{G}}{\partial \tilde{\lambda}_{c}}
= 1 - \frac{\pi}{2\Delta}
\frac{T_{K}^{2}e^{-\pi \tilde{\lambda}_{c} /\Delta} }
{\sqrt{T_K^2 e^{-\pi \tilde{\lambda}_{c} /\Delta} 
+ \sum_{\nu} \lambda_{\nu}^{2} }}.
\end{equation}
It follows from these equations that
\begin{equation}
\tilde{\lambda}_{\nu} ({\bf m},n) =
\frac{2 \Delta}{\pi}  
\frac{m_{\nu} (1-n)}{1-{\bf m}^2}
\end{equation}
and
\begin{equation}
\tilde{\lambda}_{c} ({\bf m},n) =
- \frac{2 \Delta}{\pi} 
\ln \frac{1-n}{(1-n_{d}) \sqrt{1-{\bf m}^2}} .
\end{equation}
Define a thermodynamic potential by 
\begin{eqnarray}
\tilde{\omega} ({\bf m},n) 
&\equiv&  E_{G}({\bf m},n) + n \tilde{\lambda}_{c} 
+ \sum_{\nu} m_{\nu} \tilde{\lambda}_{\nu}  
\nonumber \\
&=&
\epsilon_{d} - \mu  - 
\frac{2\Delta}{\pi} (1-n)
\nonumber \\
&& \quad 
+ (1- n) \frac{2 \Delta}{\pi} 
\ln \frac{1-n}{(1-n_d) \sqrt{1-{\bf m}^2}}  .
\end{eqnarray}
It is easy to confirm that
$\partial \tilde{\omega} ({\bf m},n) / 
\partial m_\nu = \tilde{\lambda}_\nu ({\bf m},n)$ 
and
$\partial \tilde{\omega} ({\bf m},n) / 
\partial n  = \tilde{\lambda}_c ({\bf m},n)$.

The case when the electron number is smaller than unity is
examined so far. The case where the electron number is larger
than unity can also be treated within this theoretical framework 
if the hole picture is taken.

In the presence of actual external fields,
$\Delta \mu$ and ${\bf H}$, the thermodynamic potential to be
minimized as a function of ${\bf m}$ and $n$ is given by
\begin{eqnarray}
\tilde{\Omega}_{A} 
&=& 
\tilde{\omega} ({\bf m},n) - n\Delta \mu -
\frac{1}{2}g\mu_B ({\bf m} \cdot {\bf H}) 
\nonumber \\
&=& \epsilon_{d} - \mu - T_{K}  - 
n\Delta \mu - \frac{1}{2}g\mu_B ({\bf m} \cdot {\bf H}) 
\nonumber \\
&& 
+ \frac{1}{2 \tilde{\chi}_{s}(0)} \left[
{\bf m}^{2} + \frac{1}{2} {\bf m}^{4} + \cdots \right]
\nonumber \\
&& 
+ \frac{1}{2 \tilde{\chi}_{c}(0)}  \Delta n^{2}
- \tilde{g}  {\bf m}^{2} \Delta n  + \cdots  ,
\end{eqnarray}
with $\Delta n= n-n_d$,
$\tilde{\chi}_{s}(0) = (\pi/2\Delta)/|1 - n_d|$,
$\tilde{\chi}_{c}(0) = (\pi/2\Delta) |1 - n_d|$,
and
\begin{equation}\label{Eq-g}
\tilde{g} = 
\left\{\begin{array}{cc}
\Delta /\pi, &   n_d <1
\vspace{0.2cm}\\ 
- \Delta /\pi, &  n_d >1
\end{array} \right. .
\end{equation}
Here, the results for $n_d>1$ in the hole picture are
interpreted into those in the electron picture. Then, it follows
that
\begin{equation}\label{EqBNu4}
\tilde{b}_{\nu^{4}} = \frac{2}{\tilde{\chi}_s(0)}
=\frac{4\Delta}{\pi} |1-n_{d}| = 2 k_B T_{K} .
\end{equation}
When the Friedel sum rule\cite{Shiba} is made use of, the
density of states at the chemical potential is given by
$\rho(0)  = \sin^2
\!\left(\pi n_d/2\right) /\pi \Delta$. 
When the Fermi-liquid relations,
$\tilde{\chi}_s(0)=\tilde{\phi}_{s} \rho(0)$ and
$\tilde{\chi}_c(0)=\tilde{\phi}_{c} \rho(0)$,
are made use of, it follows that
\begin{equation}\label{EqTilPhiS}
\tilde{\phi}_{s} = \frac{1}{|1-n_{d}|} 
\frac{\displaystyle \left(\pi n_d/2 \right)^2}
{\displaystyle \sin^2 \!\left(\pi n_d/2 \right)},
\end{equation}
and
\begin{equation}\label{EqTilPhiC}
\tilde{\phi}_{c} = |1-n_{d}|
\frac{\displaystyle \left(\pi n_d/2 \right)^2}
{\displaystyle \sin^2 \!\left(\pi n_d/2 \right)}.
\end{equation}


\begin{references}
\bibitem{ComThreePeak}
Photoemission electron spectra (PES) are almost the same in a
high-energy part between metallic and insulating phases in the
vicinity of the transition. These observations also imply that
the Mott-Hubbard splitting exists even in metallic phases.
\bibitem{ComNeutraility}
The density $n_d$ should be equal to that of positive ions
because of the charge neutrality of the system.
\bibitem{ComSus}
In this paper, the spin susceptibility is defined in such a way
that it does not included the factor $g^2\mu_B^2/4$.
\end{references}
\end{document}